\documentclass[preprint2]{aastex}

\usepackage{graphicx}
\DeclareGraphicsExtensions{.ps}
\shorttitle{\indent \def Dark structures in sunspot light bridges} \shortauthors{Zhang et al.}

\begin{document}

\title{Dark structures in sunspot light bridges}

\author{Jingwen Zhang\altaffilmark{1}, Hui Tian\altaffilmark{1}, Sami K. Solanki\altaffilmark{2,3}, Haimin Wang\altaffilmark{4,5,6}, Hardi Peter\altaffilmark{2}, Kwangsu Ahn\altaffilmark{5}, Yan Xu\altaffilmark{4,5,6}, Yingjie Zhu\altaffilmark{1}, Wenda Cao\altaffilmark{4,5,6}, Jiansen He\altaffilmark{1}, Linghua Wang\altaffilmark{1}}
\altaffiltext{1}{School of Earth and Space Sciences, Peking University, 100871 Beijing, China; huitian@pku.edu.cn}
\altaffiltext{2}{Max-Planck Institute for Solar System Research, Justus-von-Liebig-Weg 3, D-37077 G{\"o}ttingen, Germany }
\altaffiltext{3}{School of Space Research, Kyung Hee University, Yongin, 446-701, Republic of Korea }
\altaffiltext{4}{Space Weather Research Laboratory, New Jersey Institute of Technology, University Heights, Newark, NJ 07102-1982, USA}
\altaffiltext{5}{Big Bear Observatory, New Jersey Institute of Technology, 40386 North Shore Lane, Big Bear City, CA 92314-9672, USA}
\altaffiltext{6}{Center for Solar-Terrestrial Research, New Jersey Institute of Technology, University Heights, Newark, NJ 07102-1982, USA}

\begin{abstract}
We present unprecedented high-resolution TiO images and Fe I 1565 nm spectropolarimetric data of two light bridges taken by the 1.6-m Goode Solar Telescope at Big Bear Solar Observatory. In the first light bridge (LB1), we find striking knot-like dark structures within the central dark lane. Many dark knots show migration away from the penumbra along the light bridge. The sizes, intensity depressions and apparent speeds of their proper motion along the light bridges of 33 dark knots identified from the TiO images are mainly in the ranges of 80$\sim$200~km, 30\%$\sim$50\%, and 0.3$\sim$1.2~km~s$^{-1}$, respectively. In the second light bridge (LB2), a faint central dark lane and striking transverse intergranular lanes were observed. These intergranular lanes have sizes and intensity depressions comparable to those of the dark knots in LB1, and also migrate away from the penumbra at similar speeds. Our observations reveal that LB2 is made up of a chain of evolving convection cells, as indicated by patches of blue shift surrounded by narrow lanes of red shift. The central dark lane generally corresponds to blueshifts, supporting the previous suggestion of central dark lanes being the top parts of convection upflows. In contrast, the intergranular lanes are associated with redshifts and located at two sides of each convection cell. The magnetic fields are stronger in intergranular lanes than in the central dark lane. These results suggest that these intergranular lanes are manifestations of convergent convective downflows in the light bridge. We also provide evidence that the dark knots observed in LB1 may have a similar origin.
\end{abstract}

\keywords{Sun: sunspots---Sun: magnetic fields---Sun: photosphere---Sun: granulation}

\section{Introduction}
Sunspots, the darkest regions on the solar surface, are believed to result from strong magnetic fields inhibiting convection and thereby blocking the passage of heat from the solar interior to the surface \citep[e.g,][]{Solanki2003}. Many observations have revealed bright substructures that are associated with weaker and more horizontal magnetic fields in sunspots, such as umbral dots (UDs), light bridges (LBs) and penumbral filaments \citep[e.g.,][]
{LoughheadandBray1960,Danielson1964,BrayandLoughhead1964,BeckersandSchroter1968,BeckersandSchroter1969b,Lites1990,Leka1997,Socas2004,Jurcak2006,Riethmuller2013,Tiwari2013,Lagg2014,Felipe2017,Wang2018}. These structures are usually thought to be produced by magneto-convection in sunspots. \cite{Parker1979c} and \cite{Choudhuri1986} used the "spaghetti model" to explain the formation of UDs and penumbral grains. According to this model,  sunspots are collections of thin flux tubes loosely clustered together. Convective field-free fluid fills the gaps between the tubes. \cite{SpruitandSchramer} also interpreted penumbral filaments as being caused by intruding field-free materials from below the solar surface. However, oscillatory or overtuning magnetoconvection in the monolithic flux tube model can also explain the existence of bright substructures in sunspots \citep[e.g.,][]{KnoblochandWeiss1984,Weiss1990,Weiss1996,Rimmele1997,Rempel2009a,Rempel2009b}. In the magnetoconvection simulation of \cite{SchusslerandVolger}, no intrusion of field-free plasma into a cluster-type sunspot is needed for the production of central umbral dots. But their results can not rule out the possibility that such intrusions are relevant in the formation of light bridges.

Meanwhile, high-resolution observations have frequently revealed central dark lanes within these bright substructures in sunspots. \cite{Scharmer2002} observed dark cores inside the penumbral filaments near umbral regions. \cite{BergerandandBerdyugina2003} reported a similar central dark lane and large-scale unidirectional flows in a light bridge. Modeling efforts have also been made to understand the formation mechanisms of these dark lanes. From their realistic simulation, \cite{SchusslerandVolger} predicted that central dark lanes also exist in elongated umbral dots, which was later confirmed observationally by \cite{Bharti2007}. \cite{SchusslerandVolger} interpreted these dark lanes as the results of convective upwelling plumes accumulating plasma in the center and raising the $\tau$=1 surface. The magnetohydrodynamic (MHD) simulation of \cite{Heinemann2007} also shows that dark lanes in penumbral filaments are associated with hot upflows and weak magnetic field. This interpretation is strongly supported by the correlation between dark lanes and blue shifts in light bridges \citep{Giordano2008,Rouppe2010}, as well as the fact that dark lanes in light bridges are more prominent in radiation emitted in layers higher than the formation height of the contimuum \citep{Rimmele2008}.

Using the high-resolution TiO imaging and Fe I 1565 nm spectropolarimetric data taken by the 1.6-m Goode Solar Telescope \citep[GST,][]{Cao2010,Cao2012} at Big Bear Solar Observatory, we have studied the dark structures in two light bridges. Besides the central dark lane, we also report the detection of a new feature in the first light bridge: dot-like dark knots. While in the second light bridge, originally the central dark lane is much fainter, and some transverse intergranular lanes have been observed. However, after one day of evolution, this wide light bridge becomes narrower, and the central dark lane and dark knots are clearly detected. Here we compare the characteristics of the central dark lanes, dark dots and intergranular lanes, and discuss their possible formation mechanisms.

\section{Observations}
We mainly studied two light bridges observed with the TiO broadband filter (red continuum) and the Near InfraRed Imaging Spectropolarimeter (NIRIS) installed on GST, which is stabilized by a high-order adaptive optics (AO) system \citep{Shumko2014}. One light bridge (LB1) was observed from 17:00 UT to 22:51 UT on 2015 June 20, and the other one (LB2) was observed from 18:52 UT to 22:53 UT on 2015 June 21 and from 16:51 UT to 22:30 UT on 2015 June 22. Both were located in a sunspot in NOAA active region (AR) 12371. The observation on 2015 June 21 suffered from several interruptions and therefore we only analyzed the data acquired during the first two hours of that observation. TiO images were obtained in all these observations, whereas NIRIS data were only available on 2015 June 21 and 22. The locations of the sunspot on these three days were all very close to the solar disk center. The center of the analyzed region on 2015 June 21 has a coordinate of [x=-133$^{\prime\prime}$, y=172$^{\prime\prime}$], corresponding to a heliocentric angle of $\theta$=13.09$^o$ ($\mu$=cos $\theta$=0.974).

The TiO filter with a 10 \AA{}~bandwidth was centered at the wavelength of 7057 \AA{}. The TiO data were corrected for dark currents and flat fields. The code of the Kiepenheuer-Institute Speckle Interferometry Package (\cite{Woger2008}) was applied to produce speckle-reconstructed images with a field of view (FOV) of 77$^{\prime\prime}$$\times$77$^{\prime\prime}$. The spatial pixel size was 0$^{\prime\prime}$.0324. The cadences of the two observations were in the range of 30$\sim$60~s for the TiO data.

NIRIS employs dual Fabry-P$\acute{e}$rot etalons that provide a 80$^{\prime\prime}$$\times$80$^{\prime\prime}$ FOV and a throughput of over 90~\%. NIRIS achieves a spatial pixel size of $\sim$  0$^{\prime\prime}$.0793 at the Fe I 1565 nm line (with a bandpass of 0.25 \AA{}). The cadence of NIRIS data on the 2015 June 21 observation was 75 s. The Fe I 1565 nm line offers a high Lande g-factor of 3, which, along with its long wavelength, can help enhance signals from weak magnetic fields in the magnetograms. We inverted the NIRIS Stokes profiles using the Milne-Eddington technique, through which several key physical parameters (including the magnetic field strength, inclination angle of the field and Doppler velocity) were extracted \citep{Lites2007}. The sensitivity of the measured line-of-sight (LOS) component of the magnetic field is 10 G, i.e. the noise in the data is sufficiently low so that fields as weak as 10 G can be detected. For wavelength calibration, we set the average Doppler velocity in the darkest umbra to be zero.

\section{Data analysis and Results}

\begin{figure*}
\centering {\includegraphics[width=\textwidth]{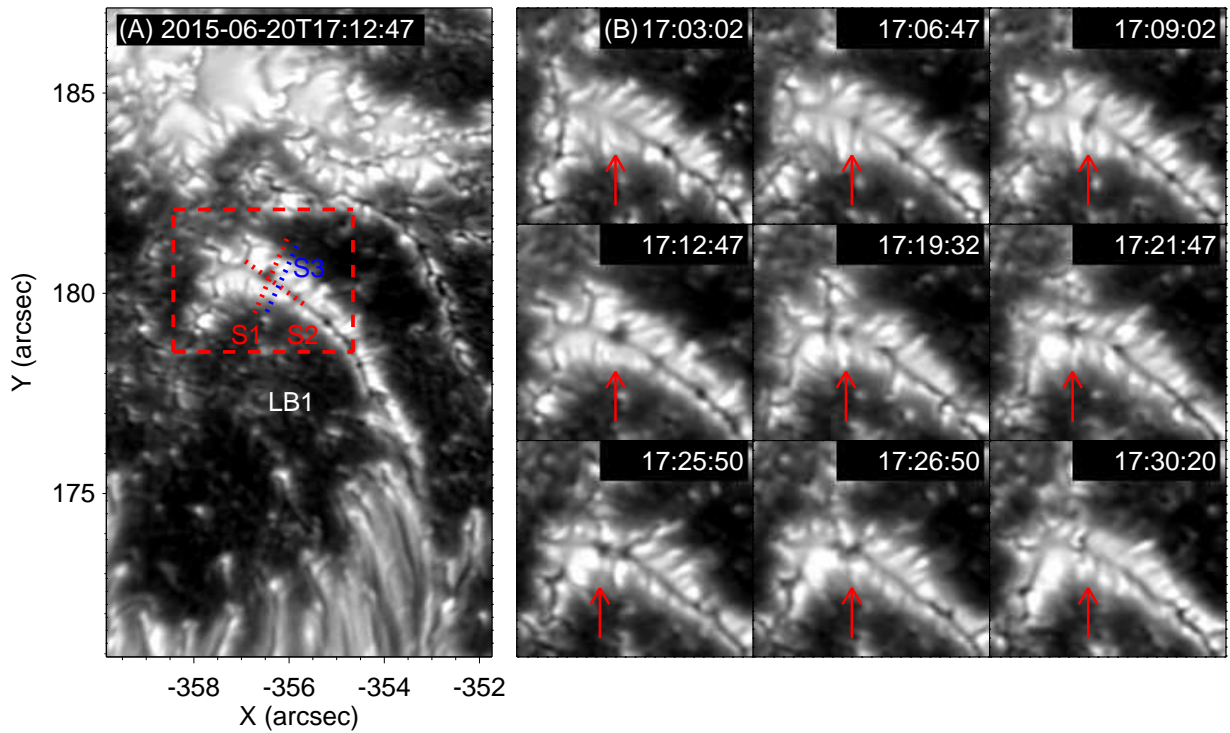}} \caption{ (A) TiO image of NOAA AR 12371 taken at 17:12:47 UT on 2015 June 20. A narrow light bridge, LB1, is located in the center of the image. The dashed red box outlines the region shown in (B). The red dotted lines S1 and S2 mark two cuts for which we produce space-time diagrams shown in Figure~\ref{fig.2} (A) and (B), respectively. The blue dotted line S3 is perpendicular to the central dark lane. (B) Temporal evolution of a dark knot (indicated by the red arrows) observed between 17:03:02 UT and 17:30:20 UT. An associated animation (m1.mpg) showing the TiO image sequence is available online.} \label{fig.1}
\end{figure*}

\begin{figure*}
\centering {\includegraphics[width=\textwidth]{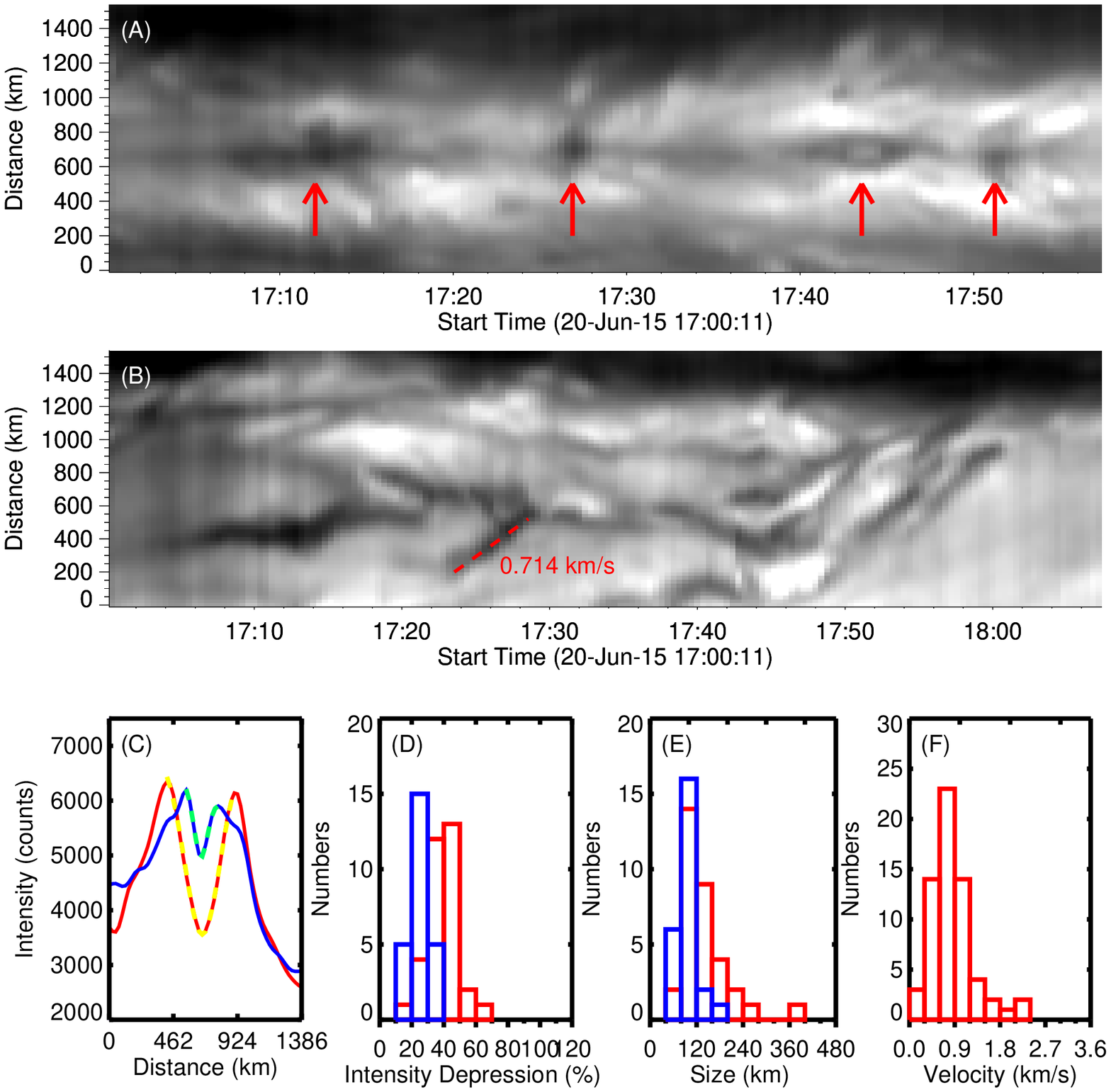}} \caption{(A) Part of the space-time diagram along cut S1 shown in Figure~\ref{fig.1}(A). The red arrows mark the crossing of cut S1 by several dark knots. (B) Part of the space-time diagram along cut S2 shown in Figure~\ref{fig.1} (A). The dashed line indicates the trajectory of a dark knot as an example, from which the apparent speeds of the  proper-motions speed can be estimated. (C) The red and blue solid lines show the TiO intensity along cuts S1 and S3, respectively. The dashed lines are single Gaussian fits to the central depressions of these curves. (D) \& (E) Histograms of intensity depression and size. The red and blue colors are for the dark knots and the central dark lane, respectively. (F) Histogram of dark knot velocity estimated from the slopes of the dark stripes exemplified in (B).} \label{fig.2}
\end{figure*}

Figure~\ref{fig.1}(A) shows a TiO image taken on 2015 June 20. The narrow light bridge (LB1) in the center of the image appears to have developed from the intrusion of a penumbral filament. The width of LB1 is $\sim$920~km. We observed a central dark lane within this light bridge. A notable feature in our observation is that several dark knot-like structures are superimposed on the dark lane, just like beads strung on a string. It is worth mentioning that similar dark knots are also visible in the images shown in \cite{Song2017} and \cite{Schlichenmaier2016}, though these features were not mentioned or discussed in their papers. The associated online animation (m1.mpg) reveals a general migration of these dark knots away from the penumbra along LB1 during the entire observation period. Figure~\ref{fig.1}(B) presents the temporal evolution of one dark knot, which appears after 17:03:02 UT and reaches its maximum size at 17:12:47 UT while moving towards east end of LB1. The large dark knot moves towards the east and merges with another dark knot at 17:21:47 UT. Afterwards, this merged dark knot encounters another eastward moving dark knot around 17:26:50 UT before disappearing. It is worth mentioning that some dark knots appear elongated in the direction perpendicular to the light bridge, somewhat similar to the transverse intergranular lanes that will be discussed later.

Figures~\ref{fig.2}(A) and (B) present space-time diagrams for cuts S1 and S2 shown in Figure~\ref{fig.1}(A), respectively. Figure~\ref{fig.2}(A) clearly reveals the crossing of cut S1 by several dark knots. Continuous sideward motion away from the central dark lane is also visible from this diagram. Along the cuts S1 and S3 shown in Figure~\ref{fig.1}(A), we can obtain intensity profiles across a dark knot and the central dark lane, respectively. Figure~\ref{fig.2}(C) shows these two intensity profiles and the single Gaussian fits to the central parts of the curves.  Through the fitting we can derive the full width at half maximum (FWHM). We defined the size of the dark knot to be the FWHM, as deduced above. Similarly, the width of the central dark lane was also taken to be the FWHM. The maximum and minimum intensities were taken directly from the central parts of the observed intensity profiles. We then calculated the intensity depression through dividing the difference between the maximum and minimum intensities by the maximum intensity. During the entire observation period, we have identified 33 dark knots and performed a similar Gaussian fitting for each of them. The red histograms in Figure~\ref{fig.2}(D) and (E) represent the distributions of size and intensity depression for these dark knots. As a comparison, we also put a cut across the dark lane at a location near each dark knot, and calculated the size and intensity depression for the central dark lane. The blue histograms in Figure~\ref{fig.2}(D) and (E) show the corresponding results. We find that the intensity depression of the central dark lane (mostly 20\%$\sim$30\%) is generally smaller than that of dark knots (mostly 30\%$\sim$50\%). The difference between the sizes of dark knots (80$\sim$200~km) and central dark lane (40$\sim$120~km) appears to be less significant, though some dark knots have notably larger sizes of 200$\sim$400~km. The dark stripes in Figure~\ref{fig.2}(B) represent the migrating trajectories of dark knots along LB1. From both the online animation (m1.mpg) and Figure~\ref{fig.2}(B), we can see a general trend that dark knots move away from the penumbra along LB1. The apparent speeds of dark knots can be estimated from the slopes of the dark stripes exemplified in Figure~\ref{fig.2}(B). We have produced 12 space-time diagrams at different positions along LB1, and identified 60 well-defined and well-isolated trajectories to calculate the velocities. The histogram presented in Figure~\ref{fig.2}(F) shows that the apparent speeds of the  proper motions of dark knots are mostly in the range of 0.3$\sim$1.2~km~s$^{-1}$. The motions of some dark knots appear to be very complex: they first move, then stop, and then move again or even move backwards slightly.

\begin{figure*}
\centering {\includegraphics[width=\textwidth]{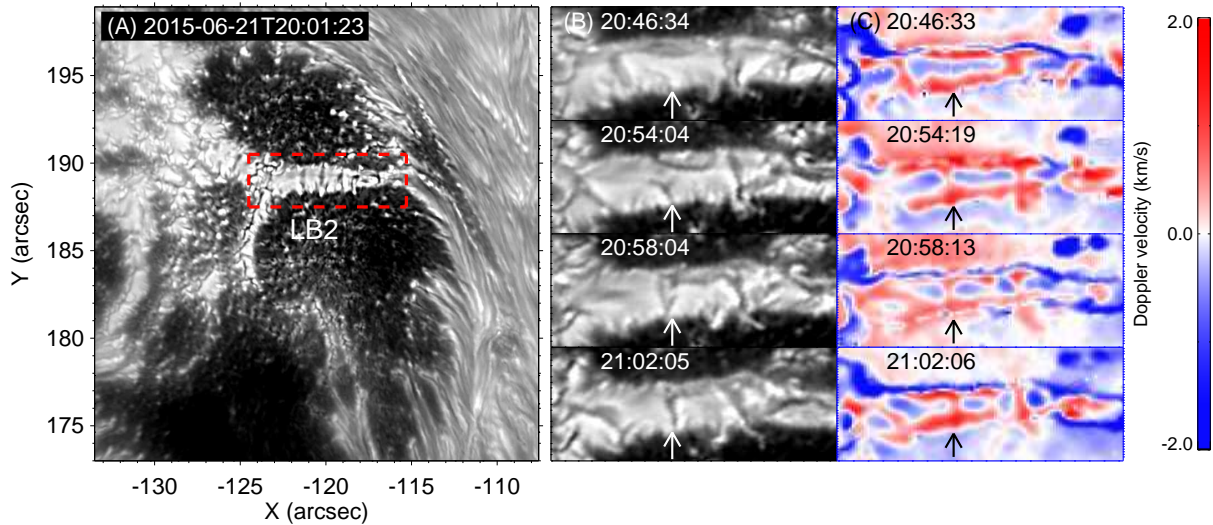}} \caption{(A) TiO image of NOAA AR 12371 taken at 20:01:23 UT on 2015 June 21. The red dashed box outlines the region shown in panels (B) \& (C) and Figure~\ref{fig.4}. (B) Image sequence of TiO showing the appearance and evolution of a transverse intergranular lane. (C) Similar to (B) but showing the Dopplergrams. Positive values (red color) represent downflows. The arrows in (B) \& (C) indicate the location of this intergranular lane.} \label{fig.3}
\end{figure*}

Figure~\ref{fig.3}(A) shows a TiO image of the same sunspot on 2015 June 21. The seeing was not as good as that on the previous day. We are mainly interested in the horizontal light bridge (LB2) in the red dashed box. The width of LB2 is $\sim$1820~km. We can see a faint central dark lane in LB2, which is much less noticeable than that in LB1. Strikingly, several intergranular lanes perpendicular to the central dark lane are seen in the image. Similar structures have also been noticed by \cite{Sobotka1994} in a light bridge, though they did not analyze these structures in detail. Our observation shows that these intergranular lanes continue to emerge in LB2, and that several of them also migrate along the light bridge away from the penumbra. Some of these intergranular lanes are not so elongated, appearing to be similar to the dark knots in LB1.

We have identified ten intergranular lanes from the TiO images taken on June 21, 2015 and obtained their intensity profiles along the lanes. We have tried to estimate their sizes and intensity depressions using the method we adopted for the dark knots observed on the previous day. It turns out that a single Gaussian fit can be applied to the intensity profiles of eight intergranular lanes. We find that their sizes (lengths) are mostly in the range of 70$\sim$420 km, and that the intensity depressions are in the range of 30\%$\sim$75\%. We have also made a space-time plot along LB2, in which we could identify paths of two intergranular lanes. The corresponding apparent proper motion speeds were estimated to be 0.69 km~s$^{-1}$ and 0.98 km~s$^{-1}$, respectively. The sizes, intensity depressions, and velocities are all comparable to those of the dark knots observed on June 20, 2015.

Figure~\ref{fig.3}(B) reveals the appearance and evolution of an intergranular lane in LB2. The simultaneously obtained Dopplergrams are presented in Figure~\ref{fig.3}(C). The growth of the intergranular lane is clearly accompanied by the appearance of redshifts. At about 20:46 UT, we can see an elongated convection cell showing a central patch of blue shift surrounded by narrow lanes of red shift. As the east boundary of the cell moves eastward, the convection cell becomes more elongated and finally splits into two after 20:54 UT, as evidenced by the appearance of a short transverse lane of red shift in the middle of the cell. In the meantime, an intergranular lane forms and grows at this location of red shift.

\begin{figure*}
\centering {\includegraphics[width=\textwidth]{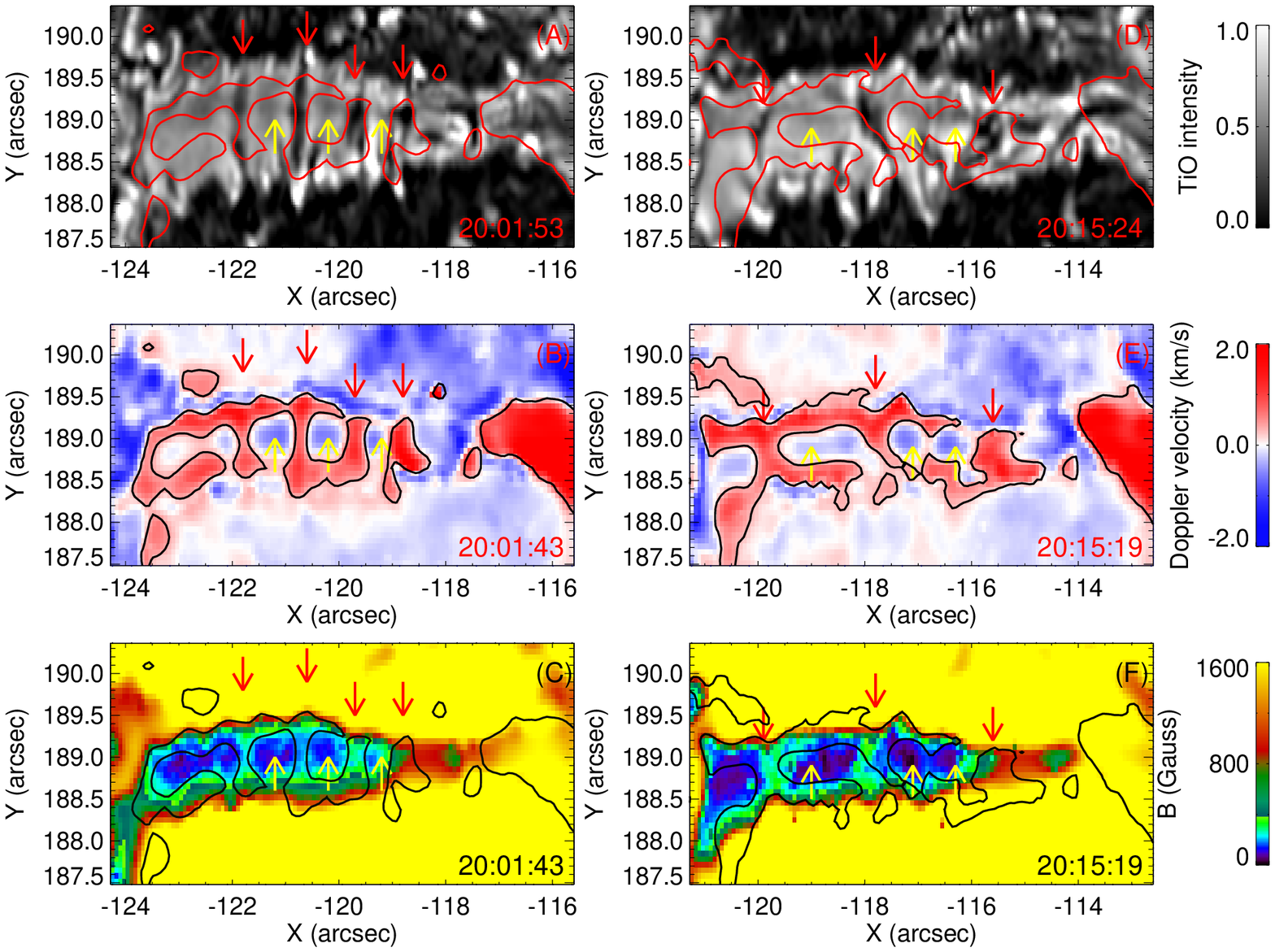}} \caption{(A)-(C) LB2 seen in images of TiO (in arbitrary units), Doppler shift and magnetic field strength around 20:01 UT on 2015 June 21. The contours indicate the Doppler velocity level of 0.1 km/s. The red arrows indicate four examples of intergranular lanes. The yellow arrows point to sections of the faint central dark lane.  (D)-(F) Same as (A)-(C) but around 20:15 UT on 2015 June 21.}\label{fig.4}
\end{figure*}

Figure~\ref{fig.4}(A)-(C) show the TiO image and simultaneously taken NIRIS data of LB2 in a smaller field of view around 20:01 on 2015 June 21. We can see at least four transverse intergranular lanes lining up in LB2. They appear to be elongated in the direction perpendicular to the light bridge, though at least one of them might also be classified as a dot-like dark knot. A faint dark lane is also visible near the center of the light bridge. The Dopplergram shown in Figure~\ref{fig.4}(B) reveals several convection cells that are characterized by patches of blue shift surrounded by narrow lanes of red shift. A comparison between Figure~\ref{fig.4}(A) \& (B) shows that the intergranular lanes correspond to redshifted regions at the two endpoints of convection cells. Whereas sections of the faint central dark lane connecting different intergranular lanes correspond to blueshifts in the center of convection cells. The spatial distribution of the magnetic field strength is presented in Figure~\ref{fig.4}(C). In agreement with previous observations of granular light bridges, LB2 has much weaker magnetic fields compared to the surrounding umbral region. The magnetic fields at the boundaries of convection cells are stronger than those in the cell interiors, which means that intergranular lanes have a relatively higher magnetic field strength. Figure~\ref{fig.4}(D)-(E) show the same images taken at a different time, reveal also similar results. At this time the three marked intergranular lanes appear to be not so elongated. Figure~\ref{fig.4}(E) \& (F) show that they also correspond to redshifts at the boundaries of convection cells and are associated with stronger magnetic fields.

\begin{figure*}
\centering {\includegraphics[width=0.8\textwidth]{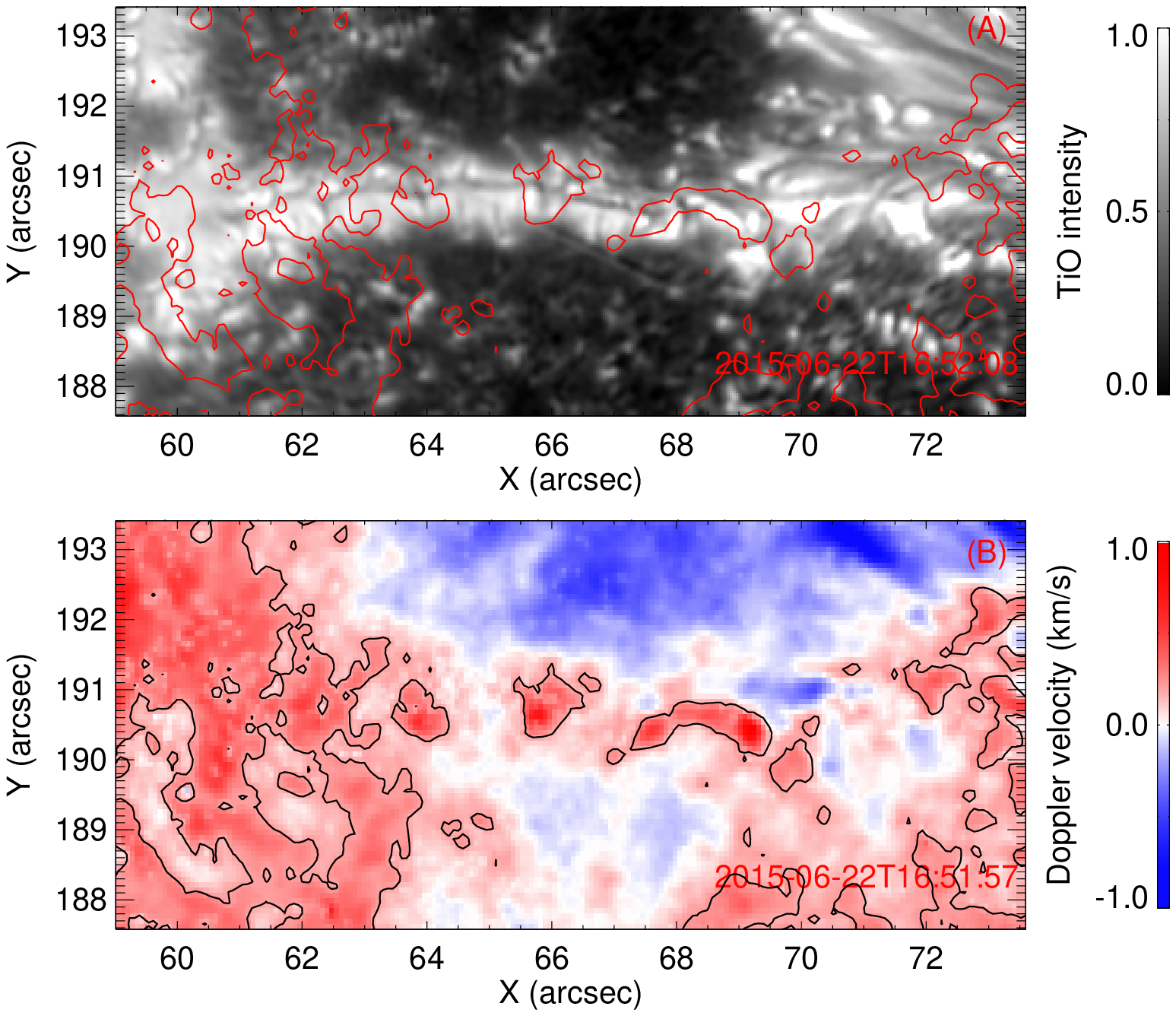}} \caption{ (A) TiO image of LB2 taken at 16:52:08 UT on 2015 June 22. (B) Dopplergram of the same region taken at 16:51:57 UT on 2015 June 22. The contours indicate the Doppler velocity level of 0.21~km~s$^{-1}$.}\label{fig.5}
\end{figure*}

Figure~\ref{fig.5} presents the TiO image and simultaneously taken Dopplergram of LB2 around 16:52:08 on 2015 June 22. After evolving for a day, the dark umbrae on both sides of LB2 lie closer to each other and the width of LB2 decreases to $\sim$730~km. Interestingly, we can see clear dark knots and a central dark lane near the center of LB2 that are similar to those in LB1. It is noticeable that these dark knots often correspond to redshifted regions in the Dopplergram. However, the Dopplergram reveals no obvious pattern of convection, which might be related to the relatively poor seeing and the fact that LB2 becomes narrower on 2015 June 22. Alternatively, the convection in a narrow LB may have a different structure than in granulation, with isolated narrow downflows separated (and possibly surrounded) by broader upflows. Higher resolution data are needed to distinguish between these different scenarios.

\section{Discussion}

\begin{figure*}
\centering {\includegraphics[width=0.8\textwidth]{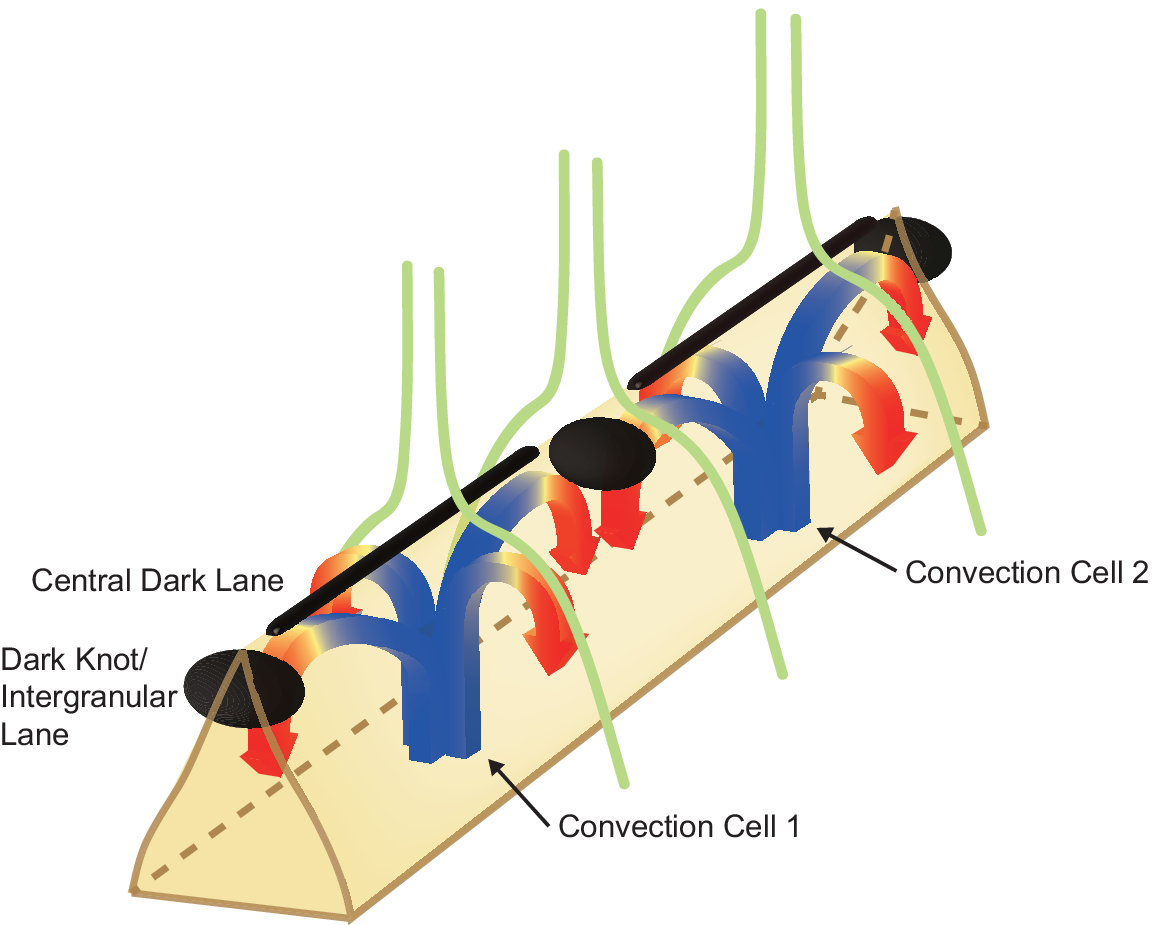}} \caption{ A cartoon showing multi-cell convection in the light bridge, which explains the formation of the central dark lane (black lines) and dark knots/intergranular lanes (black ellipses). The blue and red colors represent convective upflows and downflows, respectively. The green lines are representative magnetic field lines extending from the surrounding umbral regions. }\label{fig.6}
\end{figure*}

The central dark lanes, which are common structures in umbral dots, light bridges and penumbral filaments, have been studied extensively in the past two decades. Previous simulations and observations suggest that dark lanes are associated with convective upflows in sunspots. To explain dark lanes in umbral dots, \cite{SchusslerandVolger} came up with the following scenario: The upwelling plumes, which host weaker fields compared to the surroundings, lose their buoyancy and pile up above the solar surface, presenting cusp-like shapes in their top parts, where the strong surrounding fields merge together again. As a result of the enhanced density, the opacity in the central part gets larger and the local $\tau$=1 surface rises up to higher layers where the temperature is lower, which leads to a central dark lane in photospheric continuum images. However, only individual umbral dots are considered in this scenario. A light bridge often consists of multiple convection cells, and thus the formation of dark lanes in light bridges might be more complex than in umbral dots. Using the largest solar telescope in the world, we found different types of dark structures in light bridges: central dark lanes, dark knots and transverse intergranular lanes.

In the observation on 2015 June 20, we can see that the major dark structure running through LB1 is made up of discrete dark knots and sections of the central dark lane in between. Figure~\ref{fig.2}(C)-(E) show a comparison between the central dark lane and dark knots. Dark knots generally have a larger intensity depression. As exemplified in Figure~\ref{fig.2}(C), the intensities of dark knots are usually lower, indicating that the plasma in dark knots is likely cooler than that in the central dark lane. However, the subtle difference in intensity depression may also be partially due to the scattered light, as the larger dark knots may suffer less from the scattered light.

A striking characteristic of many dark knots in this observation is their nearly unidirectional migration away from the penumbra. This apparent motion persists for the entire observation period. We also noticed that dark knots often reveal a very dynamic evolution along the light bridge. Merging of two dark knots was occasionally observed. And some dark knots do not move at a constant speed, being nearly stationary before experiencing a sudden acceleration or even moving slightly backwards. The apparent velocity of the predominant proper motion away from the penumbra is found to be in the range of 0.3$\sim$1.2 ~km~s$^{-1}$, which appears to be comparable to the average speed (900~m~s$^{-1}$) of the bright grains on both sides of a dark lane in a light bridge analyzed by \cite{BergerandandBerdyugina2003}. The measured velocities are also similar to the velocities of inward moving penumbral grains and umbral dots, which are several hundred m~s$^{-1}$ \citep{Zhang2007,Riethmuller2008}. \cite{Kasukawa2007} reported the formation process of a light bridge and found that umbral dots rapidly split from the leading edge of penumbral filaments and migrate towards the umbra at a speed up to 1$\sim$2 ~km~s$^{-1}$. The similar velocities might suggest that these apparent motions are caused by a similar process. Penumbral grains are now considered to be the heads of penumbral filaments, which are cells of overturning magnetoconvection \citep{Rempel2009a,Rempel2009b,Rempel2011,Scharmer2011,Joshi2011,Tiwari2013}. The grains are locations of upflows and their inward movement is likely related to the inward movement of convection cells. It is possible that the migration of dark knots is also associated with the movement of convection cells.

In the observation on 2015 June 21, the central dark lane in LB2 is much fainter. There are not many knot-like dark structures in LB2. Instead, we see frequent appearance of striking transverse intergranular lanes in this light bridge. In the Dopplergrams, a chain of convective cells are clearly revealed. These intergranular lanes appear to be cospatial with convective downflows located at two boundaries of convection cells. In contrasts to this, sections of the faint central dark lane bounded by two intergranular lanes correspond to blueshifts in the center of convection cells. \cite{Lagg2014} also presented a chain of convective cells in granular light bridges. The similarities between the light bridge and quiet-sun granules led them to conclude that granular light bridges are rooted deeper in the underlying convection zone. They also found a difference in the geometry of light bridge and quiet sun granules: the light bridge granules shrink with height. It is likely that the convective divergence is weakened by the strong umbral magnetic fields on both sides, and mainly proceeds along the narrow light bridge. The reduced divergent efficiency in the direction perpendicular to a narrow light bridge may partly explain why the upwelling plasma accumulates on the solar surface and produces the central dark lane. One of factors influencing the divergent efficiency of convection cells might be the width of light bridges. In a wider light bridge the plasma can diverge and fall back more easily and efficiently, resulting in less accumulation of upwelling material and the weakening of the central dark lane. The widths of LB1 and LB2 have been measured as $\sim$ 920 ~km and $\sim$ 1820 ~km, respectively. Hence, the fact that the central dark lane in LB2 observed on 2015 June 21 is much fainter than that in LB1 may be related to the larger width of LB2. On 2015 June 22, the width of LB2 decreases to $\sim$ 730 ~km. As a result, the central dark lane in LB2 becomes much more noticeable. In addition, we find that the magnetic fields of intergranular lanes are stronger than those of central dark lanes, which might be the consequence of convective divergent flows carrying magnetic fields from the cell interiors to their boundaries.

We speculate that the dark knots observed in LB1 and the intergranular lanes in LB2 may have a similar origin. First, they have similar sizes and intensity depressions in the TiO images. Second, they both show apparent motions generally away from the penumbrae at similar speeds. Third, dark knots revealing signatures of red shift have been observed in LB2 on 2015 June 22, though the convective patterns can not be observed on that day (shown in Figure~\ref{fig.5}). It is true that the intergranular lanes in LB2 are mostly elongated in shape, whereas the dark knots in LB1 are generally roundish. However, some intergranular lanes in LB2 are also found to be nearly roundish or not so elongated at some times. Likewise, the dark knots in LB1 are not always roundish either. At some occasions some dark knots could also be classified as intergranular lanes. Perhaps, the shape of these dark structures depends on how close two convection cells are. If two adjacent convection cells are too close, they may squeeze the roundish dark knot between the two cells to form a more elongated intergranular lane. Another possibility is that the shape of these dark structures is determined by the width of light bridges. In a wider light bridge, convective downflows can occupy longer distances across the light bridge, and thus appear more elongated. A third possibility is that the dark knots are the intersections of central and intergranular lanes. However, considering the different directions of flows in the two types of dark lanes, it is unclear whether this scenario can explain the obvious red shifts at the location of dark knots or not. 

Our observations show the existence of different types of dark structures in light bridges. These dark structures all seem to be related to the convection: the central dark lane is associated with convective upflows in the center of convective cells, whereas dark knots and intergranular lanes correspond to downflows at two sides of the cells, although for the dark knots we may not rule out the possibility that they are localized downflows surrounding by upflows, which would imply a very different geometry of convection than seen in granulation.

Based on these observational results, we propose the following scenario (shown in Figure~\ref{fig.6}): Vigorous convection occurs below the light bridge and a chain of convection cells are formed. In the center of each convection cell, hot materials move upward from the convection zone, pile up in the photospheric surface and elevate the $\tau$=1 surface. As a result, a central dark lane appears on top of the light bridge. In the meantime, cool materials descends at the boundaries of the convection cells. Downflows between two adjacent convection cells lead to the appearance of dark knots/intergranular lanes and accumulation of magnetic flux.

The dynamics of the dark knots and intergranular lanes may provide insight into the evolution of convective cells in light bridges. For instance, sometimes we see merging of two dark knots, as exemplified in Figure~\ref{fig.1} (B). Since the two dark knots may correspond to the downflows at the two endpoints of a convection cell, the merging should be a reflection of the fact that this convection cell dies. The emerging of a intergranular lane shown in Figure~\ref{fig.3}(B) \& (C), however, reveals another type of evolution of a convection cell. As the central blueshifted region expands, the upflows collapse and turn into downflows. As a result, one large convective cell is split into two smaller cells. A similar process has also been frequently found in quiet-sun granules \citep[splitting or exploding granules; e.g., ][]{Ploner1999}.

In the future, more high-resolution observations and advanced numerical simulations should be performed to understand the formation and evolution of these dark structures in light bridges. In addition, it is worthwhile to investigate whether the dark knots and intergranular lanes play a role in triggering upper-atmosphere activities such as the light bridge surges or light walls \citep{Asai2001,Shimizu2009,Louis2014,Bharti2015,Yang2015,Toriumi2015a,Robustini2016,Yuan2016,Zhang2017,Hou2017,Tian2018}.

\section{Conclusions}

Using the high-resolution TiO imaging and Fe I 1565 nm spectropolarimetric data taken by the 1.6-m Goode Solar Telescope at Big Bear Solar Observatory, we have analyzed the dark structures in two light bridges of a sunspot in Active Region 12371. The main results and conclusions are summarized as follows.

First, the dark structures in LB1 mainly consist of a central dark lane. This is subdivided into sections by striking dark knots. The dark knots mostly have intensity depressions of  30\%$\sim$50\% and sizes of 80$\sim$200~km. They generally migrate away from the penumbra at speeds of 0.3$\sim$1.2~km~s$^{-1}$.

Second, LB2 has a much fainter central dark lane but striking transverse intergranular lanes on 2015 June 21. The intensity depressions and lengths of these intergranular lanes are mostly in the ranges of 30\%$\sim$75\% and 70$\sim$420~km, respectively. These intergranular lanes also reveal apparent motions generally away from the penumbra at speeds similar to those of dark knots. They are cospatial with convective downflows located at two edges of convection cells. On the contrary, sections of the central dark lane bounded by two intergranular lanes correspond to blueshifts in the center of convection cells. We also find that LB2 evolves into a much narrower light bridge on 2015 June 22, when a central dark lane and dark knots can be clearly identified.  

These results suggest that a narrow light bridge is made up of a chain of convection cells, just like broader, granular light bridges. In the center of each convection cell, upflows raise up the $\tau$=1 surface and result in a central dark lane. Whereas on the two endpoints of the convection cells, depending on the width of a light bridge, converging downflows may create dot-like dark knots or elongated intergranular lanes. By observing the dynamics of these dark structures, we now can infer some detailed information about the evolution of the convection cells.

\begin{acknowledgements}
{We thank Prof. J. Chae and Dr. A. Lagg for helpful discussion. This work is supported by NSFC grants 41574166 and 11790304 (11790300), NSF grants AGS-1408703 and AGS-1539791, the Max Planck Partner Group program, the Recruitment Program of Global Experts of China, the European Research Council (ERC) under the European Union's Horizon 2020 research and innovation programme (grant agreement No. 695075), and the BK21 plus program through the National Research Foundation (NRF) funded by the Ministry of Education of Korea. BBSO operation is supported by NJIT, US NSF AGS-1821294 and AGS-1821294 grants. The GST operation is partly supported by the Korea Astronomy and Space Science Institute (KASI), Seoul National University, and the Strategic Priority Research Program of CAS with Grant No. XDB09000000. Authors thank BBSO staff for their help during the observations. }

\end{acknowledgements}

\end{document}